# Single mode lasing in transversely multi-moded PT-symmetric microring resonators

Hossein Hodaei[1], Mohammad-Ali Miri[1], Absar U. Hassan[1], William E. Hayenga[1], Matthias Heinrich[1,2], Demetrios N. Christodoulides[1], and Mercedeh Khajavikhan[1*]

[1] *CREOL, The College of Optics and Photonics, University of Central Florida, Orlando, Florida 32816-2700, USA*
[2] *Institute of Applied Physics, Abbe Center of Photonics, Friedrich-Schiller-Universität Jena, Max-Wien-Platz 1, 07743 Jena, Germany*

*Corresponding author: mercedeh@creol.ucf.edu

**Single mode lasing is experimentally demonstrated in a transversely multi-moded InP-based semiconductor microring arrangement. In this system, mode discrimination is attained by judiciously utilizing the exceptional points in a parity-time (PT) symmetric double microring configuration. The proposed scheme is versatile, robust to small fabrication errors, and enables the device to operate in a stable manner considerably above threshold while maintaining spatial and spectral purity. The results presented here pave the way towards a new class of chip-scale semiconductor lasers that utilize gain/loss contrast as a primary mechanism for mode selection.**

Integrated photonic laser systems with larger cross sections are desirable for many applications since they allow for higher energies within the cavities while managing the thermal load and keeping the impact of optical nonlinearities under control. Unfortunately, however, merely enlarging the transverse dimensions of the waveguides inevitably gives rise to competing higher-order spatial modes. This, in turn, compromises the spectral and spatial fidelity of the laser and limits the power allocated within a specific mode [1]. These limitations exist on all scales, and may even be exacerbated in chip-scale semiconductor lasers, where the large gain bandwidths of the active media already pose a challenge in promoting single-mode operation [2].

So far, utilizing intra-cavity dispersive elements has been the primary approach for longitudinal mode selection [3], while tapering along the direction of propagation, engineering the refractive index in the cross section, as well as evanescent filtering are some of the extensively explored techniques to enforce single spatial mode operation in such arrangements [4-7]. Yet, in spite of their success, they are not always compatible with on-chip microcavity structures and in most cases are quite sensitive to small fabrication tolerances. In this respect, it would be desirable to explore alternative avenues to address these issues.

Lately, the selective breaking of parity-time (PT) symmetry has been proposed as a viable strategy for obtaining single transverse mode operation [8]. It is suggested that by pairing an active resonator/waveguide with a lossy but otherwise identical partner, it is possible to enforce single-spatial-mode performance even in the presence of strong mode competition in multi-moded laser/amplifier configurations. In general, a structure is considered to be PT-symmetric if it is invariant under the simultaneous action of the $\mathcal{P}$ (space) and $\mathcal{T}$ (time) inversion operations [9]. Despite

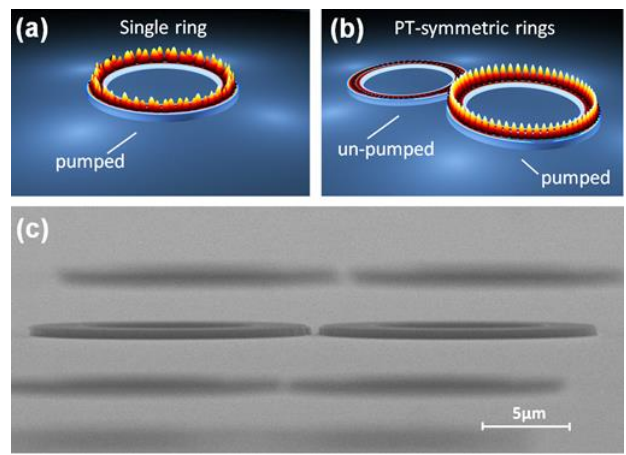

Fig. 1 (a) In a single microring resonator, multiple transverse and longitudinal modes can lase simultaneously, (b) on the other hand, in a PT-symmetric arrangement, only one longitudinal mode with the lowest order transverse profile can lase, and (c) an SEM image of a PT-symmetric double ring structure.

having a non-Hermitian representation, such a system may still support an entirely real spectrum. While originally developed in the context of quantum mechanics [9,10], such notions have recently attracted considerable attention in different areas of optics such as photonic lattices, microresonators, gratings, and lasers [11-22]. In optical settings, a structure is PT symmetric if the real part of the refractive index is an even function of position, whereas the imaginary component (representing gain and loss) exhibits an odd profile.

In a recent work, we have shown experimentally how PT symmetry can be utilized to enable *longitudinal* mode discrimination in spatially single-moded, but spectrally multi-moded microring laser arrangements without compromising the output power [22]. The principle of operation is based on the spontaneous breaking of PT symmetry, which serves as a virtual lasing threshold and encourages single longitudinal mode operation. The PT-symmetry is achieved using a coupled microring system in which the pump power was withheld from one of the cavities. In such settings, the net differential gain associated with the spectral curvature of the lineshape function is systematically enhanced- a direct byproduct of the existence of a non-Hermitian exceptional point [22]. Exceptional points have been also shown to reverse the dependence of lasing threshold on pump power [18,19].

In this letter, we show that PT symmetry can also be utilized in promoting the fundamental *transverse* mode in spatially multi-moded micro-ring lasers. In fact, as shown in [8], the virtual threshold at the exceptional point $g/\kappa = 1$ introduces an additional degree of freedom, e.g., the coupling constant $\kappa$ between the active and the lossy cavity, mediated by the evanescent overlap of their respective modes. As it is well known, higher order spatial modes systematically exhibit stronger coupling coefficients due to their lower degree of confinement. Consequently, in a PT symmetric arrangement, the fundamental mode is the first in line to break its symmetry as the gain increases (when $g > \kappa$), thus experiencing a net amplification. On the other hand, for this same gain level, the rest of the modes retain an unbroken symmetry and therefore remain entirely neutral. Indeed, following our approach, one can globally enforce single-mode behavior both in the spatial and spectral domain. The coupled mode analysis of the PT-symmetric coupled microring systems is presented in Supplementary Section 1.

Figures 1(a) and (b) schematically illustrate the transition from multimode behavior in a microring laser to single mode operation in a twin-ring configuration as enabled by preferential PT symmetry breaking. Our experiments were conducted in high-contrast active ring resonators based on quaternary InGaAsP (Indium-Gallium-Arsenide-Phosphide) multiple quantum wells embedded in $SiO_2$ (silicon dioxide) and air as shown in Fig. 1(c). Based on our measurements, we estimate the quality factor of the fabricated microrings to be on the order of 120,000 (for more information regarding the fabrication and cavity Q-factor, see the supplementary information. The gain bandwidth of the active medium spans the spectral region between 1290 and 1600 nm. Whereas the quantum wells are present in all wave-guiding sections, gain and loss is provided by selectively pumping the respective rings (pump wavelength: 1064 nm). Accordingly, the effective pump powers are proportional to the geometric overlap between the active medium and the pump profile. In our proof-of-principle design, a ring of an outer radius of 6 μm and waveguide dimensions of 0.21 μm × 1.5 μm were chosen to realize

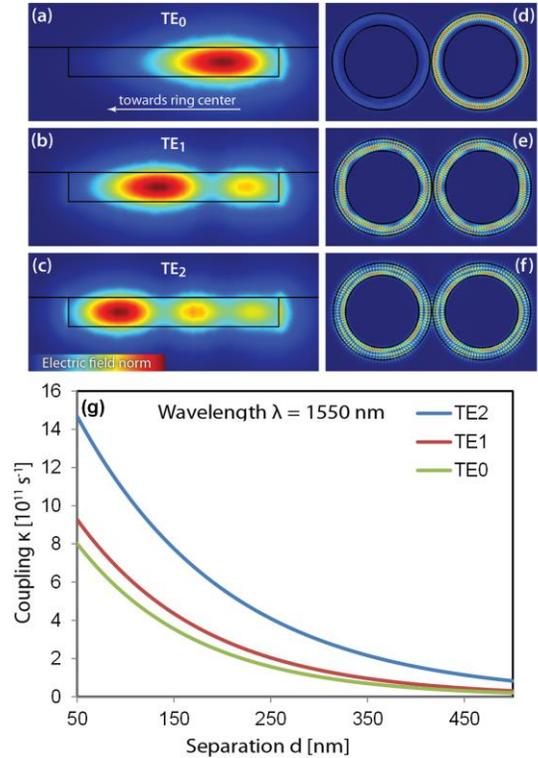

Fig. 2 (a-c) Intensity distributions in a microring resonator with a cross section $0.21\ \mu m \times 1.5\ \mu m$ and a radius of R=6 μm as obtained by finite element simulations for the first three transverse modes, (d-f) intensity distribution of these same modes within the PT symmetric ring resonators. While the $TE_0$ mode operates in the broken PT symmetry regime and lases, all other modes remain in their exact PT phase and therefore they stay neutral (below lasing threshold), and (g) Exponential decay of the temporal coupling coefficients $\kappa$ with cavity separation $d$. Higher order modes exhibit larger coupling coefficients than their lower-order counterparts.

comparably large free spectral range ($\Delta\lambda_{FSR} \sim 16$ nm) and readily discernible three sets of transverse modes (see Fig. 2).

Figures 2(a-c) depict the transverse intensity profiles of different spatial modes supported by such a single microring resonator. The curvature of the ring imposes a radial potential gradient, which deforms the mode fields into whispering-gallery-like distributions. Whereas the centroid of all modes shifts towards the ring center, the exponential decay outside the ring still grows strongly with the mode order. As shown in Figs. 2(d-f), for a certain coupling coefficient, set by the distance between the two rings, the transverse $TE_0$ field is the only mode to break its PT symmetry while all the higher order modes ($TE_1$, $TE_2$) are still in the unbroken PT phase and hence occupy both rings equally. This behavior is also evident in Fig. 2(g) where the coupling strength between different transverse modes is depicted as a function of the separation between the two rings. For a fixed distance, the coupling coefficient increases with the order of the transverse mode, since the effective indices of higher order modes lie closer to that of the surrounding medium, allowing for stronger evanescent interactions across the cladding region. This trend persists for all wavelengths, and in conjunction with the difference in confinement enables PT symmetry breaking to be employed as a mode-selective virtual loss. More information regarding the coupling factor and mode confinement is presented in Supplementary Section 2.

The emission spectra of the fabricated microring structures have been collected using a micro-PL characterization set-up as described in the Supplementary Section 3. Utilizing a rotating diffuser, the output of a single mode pump laser is converted to spatially incoherent light with a large and uniform spot size at the sample plane. The knife-edge in the path of the pump beam allows the selective withholding of illumination from specific rings. The location of the shadow in respect to the rings is adjusted by translating the knife-edge and is monitored via the incorporated confocal microscope. The intensity distribution and the spectrum of the light oscillating in the microrings are then observed via scattering centers by means of an infrared camera and a grating spectrometer, respectively.

In order to illustrate the selective breaking of PT symmetry, we consider a scenario where a coupled arrangement of identical microring resonators is evenly illuminated by the pump beam. In this case, as shown in Fig. 3(a), every resonance in each ring splits into a doublet. In our system, the modes can be distinguished both theoretically and experimentally by considering their wavelength splitting (coupling strength) as well as their corresponding free spectral ranges. For example, the $TE_1$ supermodes exhibit greater frequency splitting compared to the $TE_0$ ones due to their higher coupling coefficient. In the PT-symmetric system where only one of the two rings is pumped, the $TE_0$ modes preferentially undergo PT symmetry breaking due to their higher modal confinement and fuse into a singlet. In addition, given that different longitudinal modes experience different amounts of gain, one can restore the PT symmetry of one of the $TE_0$ modes by adjusting the pump level. As a result, global single mode operation (spectrally and spatially) can be achieved in this twin-microring system, as only one single longitudinal resonance of the fundamental $TE_0$ mode experiences sufficient gain to induce PT symmetry breaking. Figure 3(b) shows the fusion of a doublet in the frequency domain and the formation of a single lasing mode. The effect of the deviation from the perfect PT-symmetry condition due to nonlinear and thermal mismatches between the two rings is discussed in Supplementary Section 4.

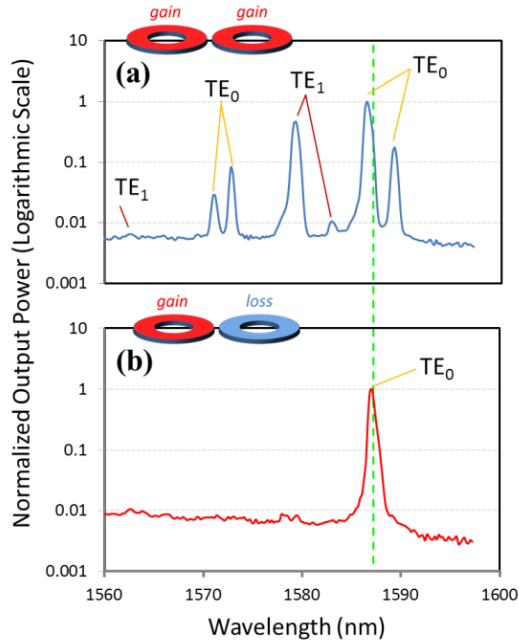

Fig. 3 PT-enforced single mode operation in the presence of higher order transverse modes. (a) Measured emission spectrum from a coupled arrangement of evenly pumped microrings, comprised of various $TE_0$ and $TE_1$ modes. (b) Global single-mode operation in the PT arrangement. Selective breaking of PT symmetry is used to suppress the entire set of $TE_1$ modes as well as competing longitudinal $TE_0$ resonances. The minimum separation between the coupled rings is 50 nm. The resolution of the spectrometer is set at ~0.5 nm. More refined measurements using scanning Fabry-Perot techniques reveal a linewidth of ~10 GHz at 1.5 times the threshold.

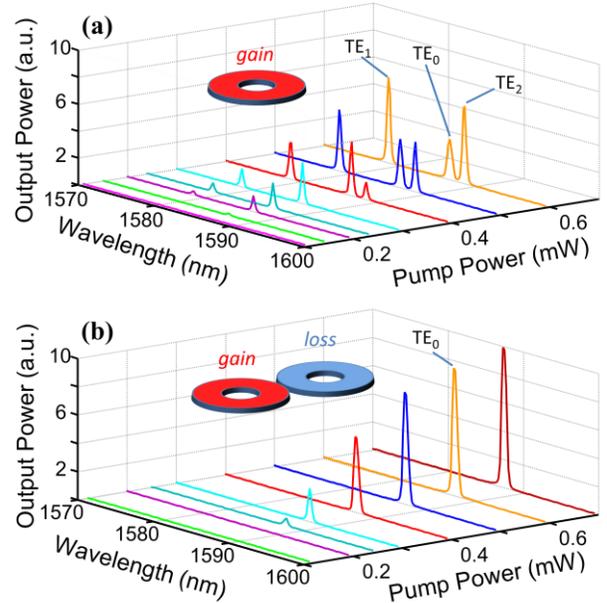

Fig. 4 Comparison between the spectral evolution of a single microring laser and the corresponding PT-symmetric arrangement as a function of the pump power. While higher order transverse modes $TE_1$ and $TE_2$ appear in the lasing spectrum of the single microring laser (a), the PT laser remains purely single-moded (b).

Figure 4 compares the modal behavior of a single ring and a PT-symmetric double ring system. Figure 4(a) shows the spectrum obtained from a single microring resonator supporting up to three transverse modes. Due to their relatively close lasing thresholds, $TE_0$, $TE_1$ and $TE_2$ simultaneously lase. Note that, in an isolated ring, a decrease of the overall pump power does in practice yield very small selectivity between these modes. On the other hand, when the active ring of the PT-symmetric double-ring arrangement is supplied with the same pump power as in Fig. 4(a), the higher order modes are readily suppressed with a fidelity of over 25 dB (Fig. 4(b)).

It is also instructive to compare the output efficiency of the PT-symmetric laser with that of a standalone microring resonator. The characteristic light-light curves shown in Fig. 5(a) demonstrate that despite a slightly higher lasing threshold, the PT arrangement exhibits approximately the same slope efficiency as the single ring. What is important, however, is that in the PT case, the entire power is now emitted into the broken-PT resonance, whereas in the single ring system it is shared among a mixture of competing modes (Fig. 5(b)). It should be noted that, the threshold in the PT system is set by the coupling between the rings and the loss of the un-pumped cavity and can be further modified through design. The technique presented here for transverse mode selection, differs from spatial mode filtering techniques that are typically used in lasers. For example, in integrated settings, higher order transverse modes can be suppressed by introducing a mode dependent loss through the evanescent coupling of the modes to a lossy waveguide. [1,24]. Although both techniques rely on the fact that various transverse modes have different confinements, in the evanescent coupling schemes, all modes including the fundamental one, experience loss, though to a varying degree. Therefore, the resulting mode suppression rendered by such methods is incremental and comes at the expense of undesirable losses in the fundamental mode. In contrast, the abrupt onset of PT symmetry breaking at the exceptional point relocates the selected mode to the active region, whereas all higher order modes remain equally distributed between loss and gain. Consequently, our approach offers a high degree of mode discrimination, without a detrimental effect on the overall lasing efficiency. Theoretical analysis comparing these two methodologies also supports this conclusion (see Supplementary Section 5). Our analysis shows that the maximum gain to operate in a single mode for the above parity-time symmetric system is ~20 times larger than that for a ring that is evanescently coupled to a waveguide in an optimal condition.

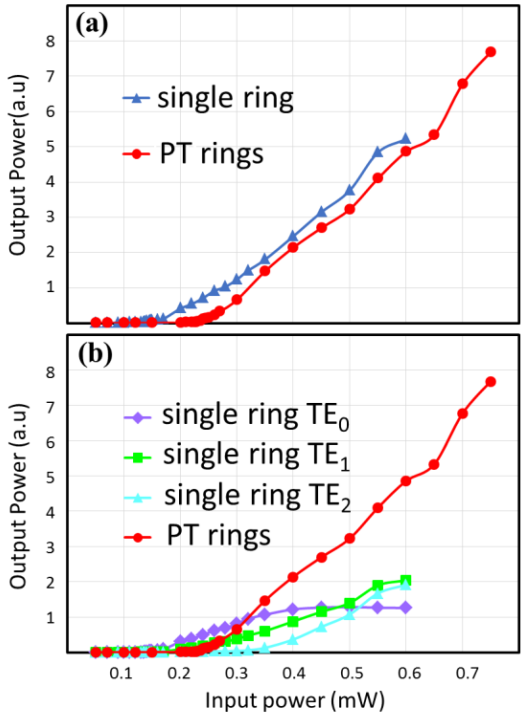

Fig. 5 Light-light characteristics. (a) Despite a slightly higher lasing threshold, the PT-symmetric arrangement exhibits the same slope efficiency, as an isolated microring. (b) The desired $TE_0$ mode of the PT system carries much higher power than any mode in the single microring. In the single ring, different transverse modes can be distinguished from their onset of lasing.

In conclusion, we have experimentally shown that single mode behavior can be effectively enforced in transversely multi-moded parity-time symmetric coupled cavities. The proposed method establishes mode selectivity through the utilization of an exceptional point and results in a significantly increased extracted power from the fundamental mode without compromising the slope efficiency. Our approach is versatile, scalable and can be applied to a wide range of on-chip laser systems.

**ACKNOWLEDGMENT**

The authors acknowledge the financial support from NSF CAREER Award (ECCS-1454531), ARO (W911NF-14-1-0543, W911NF-16-1-0013), NSF (ECCS-1128520), and AFOSR (FA9550-12-1-0148 and FA9550-14-1-0037). M.H. was supported by the German National Academy of Sciences Leopoldina (LPDS 2012-01 and LPDR 2014-03). We would also like to acknowledge M.S. Mills for helpful discussions.

# Supporting Information

**Single mode lasing in transversely multi-moded PT-symmetric microring resonators**

*Hossein Hodaei[1], Mohammad-Ali Miri[1], Absar U. Hassan[1], William E. Hayenga[1], Matthias Heinrich[1,2], Demetrios N. Christodoulides[1], and Mercedeh Khajavikhan[1,*]*

This document provides supplementary information to "Single mode lasing in transversely multi-moded PT-symmetric microring resonators". The first Section is devoted to the coupled-mode analysis of the PT microring laser system. The coupling coefficient and the confinement factor are discussed in Section 2. The third section describes the experimental setup used for the measurements of the lasing spectra. In section 4, we show that single mode operation based on PT symmetry breaking is robust with respect to perturbations and deviations from exact PT conditions. Finally, in Section 5, we compare the mode filtering based on selective PT symmetry breaking with traditional schemes based on evanescent coupling of the active device to a lossy structure.

**S1. Coupled mode analysis of the PT-symmetric coupled microring laser**

Within the framework of coupled mode theory, the temporal evolution of the modal amplitudes $a_n, b_n$ in the two cavities can be described as:

$$\frac{da_n}{dt} = -i\omega_n a_n + i\kappa_n b_n + \gamma_{a_n} a_n \quad \text{(S1a)}$$

$$\frac{db_n}{dt} = -i\omega_n b_n + i\kappa_n a_n + \gamma_{b_n} b_n \quad \text{(S1b)}$$

where the index $n$ summarizes both the transverse and longitudinal index of each mode, $\omega_n$ is the resonance frequency of the the $n^{th}$ mode, $\kappa_n$ represents the coupling coefficient between the two mode, and $\gamma_{a_n}$ and $\gamma_{b_n}$ show the modal gain or loss (depending on their sign) experienced by the $n^{th}$ mode of the two cavities. The gain or loss factors are obtained as $\gamma_{a_n,b_n} = g_{a_n,b_n} - 1/\tau_n$ where $g_{a_n,b_n}$ represents the gain in the two cavities and $1/\tau_n$ shows the intrinsic losses due to absorption, scattering, and radiation. Note that the condition of PT symmetry is satisfied when gain and loss in the two resonators are exactly balanced, i.e., $\gamma_{b_n} = -\gamma_{a_n} = \gamma_n$. Assuming eigenmode solutions of the form $(a_n, b_n) = (A_n, B_n)e^{-i\omega t}$, the two eigenfrequencies associated with the pair of $n$th supermodes are obtained to be:

$$\omega_n^{(1,2)} = \omega_n + i\left(\frac{\gamma_{a_n} + \gamma_{b_n}}{2}\right) \pm \sqrt{\kappa_n^2 - \left(\frac{\gamma_{a_n} - \gamma_{b_n}}{2}\right)^2} \quad (S2)$$

When the two cavities are equally pumped, i.e., $\gamma_{b_n} = \gamma_{a_n} = \gamma_n$, this relation reduces to:

$$\omega_n^{(1,2)} = \omega_n + i\gamma_n \pm \kappa_n. \quad (S3)$$

Evidently, in this case, both supermodes experience the same amount of gain $\gamma_n$, and can therefore simultaneously lase at two different frequencies with a splitting of $\Delta\omega = 2\kappa_n$. On the other hand, when the pump beam illuminates only one of the two cavities in a PT-symmetric manner ($\gamma_{b_n} = -\gamma_{a_n} = \gamma_n$), the frequencies of the two supermodes are obtained from:

$$\omega_n^{(1,2)} = \omega_n \pm \sqrt{\kappa_n^2 - \gamma_n^2}. \quad (S4)$$

Note that, in this case, two different regimes can be identified. As long as the gain loss contrast $\gamma_n$ remains below the coupling coefficient $\kappa_n$ between a pair of modes, the corresponding supermodes remain neutral (exact PT phase regime). However, once the gain/loss contrast exceeds coupling, a complex conjugate pair of amplifying and attenuating modes form (broken PT symmetry regime). The boundary between these two regimes is determined by the PT symmetry breaking threshold $\gamma_n = \kappa_n$ and this is the key idea behind the mode selection property of a PT-symmetric microring laser. The PT symmetry breaking point is an exceptional point in which not only the eigenvalues but also the eigenvectors become identical. Such exceptional points arise in non-Hermitian or open systems.

### S2. Coupling coefficient and confinement factor

As discussed in the main text, the coupling coefficient and the confinement factor (related to the overlap of the field with active regions and therefore estimates the modal gain) play a crucial role in determining the lasing or neutral oscillation of a mode. Here we present additional information about the wavelength dependency of these coefficients. The coupling coefficients between different transverse modes are calculated from finite element simulations for different separation between the two cavities. At a wavelength $\lambda_0 = 1550$ nm, the results are shown in Fig. 2(g) of the main text. It should be noted however that coupling coefficient also varies with frequency. As a result, different longitudinal modes experience different coupling coefficients. This is in agreement with the experimental results presented in Fig. 3 (a) in the main text.

$$\Gamma = \frac{\int_{V_a} dr |<S(r)>|^2}{\int_{-\infty}^{+\infty} dr |<S(r)>|^2}, \quad (S5)$$

where $|<S(r)>|$ is the time averaged Poynting vector and the integration in the nominator is taken over the cross section of the microring resonator. As the figure clearly indicates, higher order transverse modes exhibit lower confinement and therefore experience less modal again. This behavior is again in favor of selective PT symmetry breaking for the fundamental transverse mode.

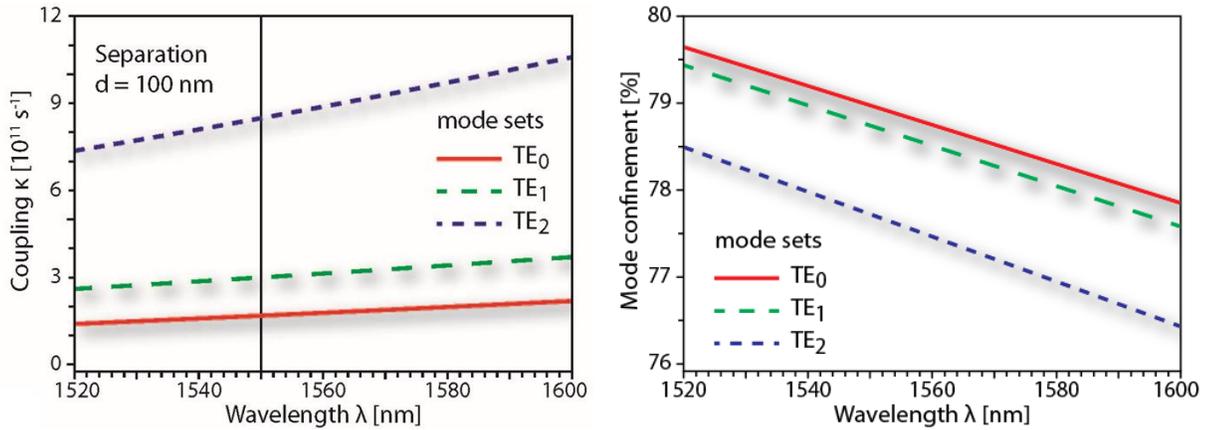

Figure S1. The coupling coefficient and the confinement factor. Provided for different transverse modes of a microring resonator with respect to wavelength when the smallest distance between the rings is 100 nm.

**S3. Experimental setup**

Figure S2 schematically shows the experimental set-up used for characterization of the PT laser. After passing through a rotating diffuser, the pump beam (1064 nm pulsed laser with a pulse duration of 15 ns and a repetition rate of 290 kHz) is projected to the sample via an NIR objective with a numerical aperture of 0.42. The shaped pump beam can therefore uniformly illuminate a circular area of 40 μm diameter on the sample plane. In order to selectively block certain areas of the sample from pump illumination we utilize a knife-edge as shown in Fig. S2. The location of the knife-edge is controlled by a translation stage while the location of its shadow on the sample plane is monitored via a confocal microscope. In this manner, different scenarios of evenly pumped and PT-symmetric coupled microring resonators can be realized (see Fig. S3).

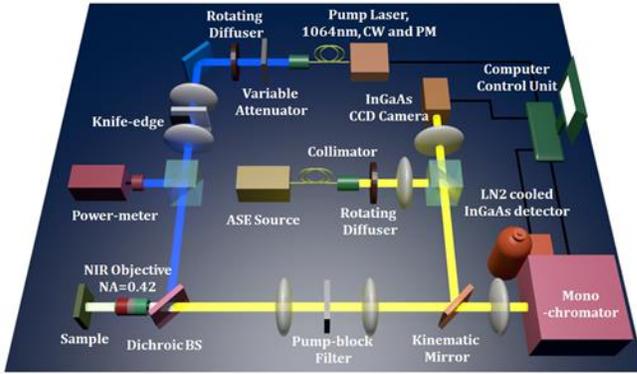

Figure S2. A Schematic of Micro-photoluminescence characterization set-up. A knife-edge is used to spatially shape the pump beam and selectively illuminate the microrings.

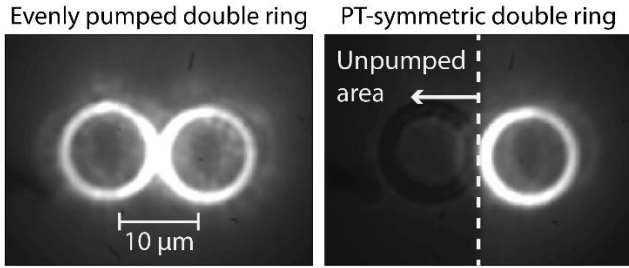

Figure S3. Realization of a PT-symmetric microring resonator pair. A knife edge was used to selectively withhold the pump beam from one of the rings.

## S4. Perturbations from exact PT symmetry conditions

As mentioned in previous section, in general, PT symmetry demands a delicate balance between the gain and loss in the microring resonators. However, such conditions may not be fulfilled in practice. This can happen due to various uncontrolled loss mechanisms, pump inhomogeneities, and the interdependency of the gain/loss and the refractive index as imposed by Kramers-Kronig relations. In addition, even if the condition of PT symmetry is fulfilled for a specific pair of modes, such requirement may not be exactly satisfied for other modes. The question naturally arises as to whether deviations from exact PT symmetry can affect the single mode operation of the PT microring laser.

Consider again Eq. (S3) which shows the eigenfrequencies of two coupled microring resonators with general gain/loss coefficients. In order to ensure single mode lasing operation in such system, one should find a coupling coefficient such that $\kappa_n < |(\gamma_{a_n} - \gamma_{b_n})/2|$ for the fundamental mode while keeping $\kappa_n < |(\gamma_{a_n} - \gamma_{b_n})/2|$ for the rest of the mode pairs. On the other hand, according to Eq. (3) each pair of modes exhibit an offset amplification or attenuation term given by $(\gamma_{a_n} + \gamma_{b_n})/2$. To prevent any amplification of the undesired modes, one can therefore enforce the condition of $\gamma_{b_n} < -\gamma_{a_n}$. This means that the amount of loss in the second cavity should be slightly higher than the amount of gain in the main cavity.

## S5. Comparison of mode filtering based on PT symmetry breaking and evanescent coupling

As mentioned in the main text, the transverse mode selection effect based on selective PT-symmetry-breaking is by nature different from the evanescent mode filtering technique which has been used for the suppression of higher order modes in semiconductor lasers (see Fig. S4). In order to show the difference between the two schemes, here, we compare the modal gain discrimination between the fundamental transverse mode $TE_0$ and its closest competing mode $TE_1$. In a PT-symmetric arrangement, we denote the coupling between these modes and their associated modes in the secondary cavity by $\kappa_0$ and $\kappa_1$ respectively. According to Eq. (S4), the gain of these two modes in the PT-symmetric arrangement is given by $\sqrt{g^2 - \kappa_0^2}$ and $\sqrt{g^2 - \kappa_1^2}$ respectively. As a result, the maximum achievable gain for the fundamental mode, while preventing the second mode from lasing, is given by

$$g_{\max,\text{PT}} = \sqrt{\kappa_1^2 - \kappa_0^2} \quad (S6)$$

Next, consider the case where an identical active microring is coupled to a waveguide with similar transverse geometry. In this case, assuming that the bus waveguide is not externally excited, coupled mode equations can be written as [S1]:

$$\frac{da_n}{dt} = \left(-i\omega_n + g_n - \gamma_{e_n}\right)a_n \quad (S7a)$$

$$s_t = i\mu_n a_n \quad (S7b)$$

where $a_n$ represent the modal amplitude of the $n^{th}$ mode in the cavity, $\gamma_n$ shows the net gain/loss (depending on its sign) in an isolated cavity, $\gamma_{e_n}$ represents a loss due to external coupling to the bus waveguide, $|s_t|^2$ is the output power and finally $\mu_n$ shows the coupling between the cavity and the waveguide. Note that $\gamma_{e_n}$ and $\mu_n$ are not independent parameters and indeed one can show that $\gamma_{e_n} = \mu_n^2/2 = \pi n_{g_n} \kappa_n^2 r/c$ where $n_{g_n}$ represents the group index of the $n^{th}$ transverse mode, $r$ represents the radius of the microring and $c$ is the velocity of light. Under these conditions the total gain experienced by each mode is given by $g_n - \pi n_{g_n} \kappa_n^2 r/c$. As a result, in this case the maximum achievable gain for single mode operation is

$$g_{\max,\text{EC}} = \pi r \left(n_{g_1}\kappa_1^2 - n_{g_0}\kappa_0^2\right)/c \quad (S8)$$

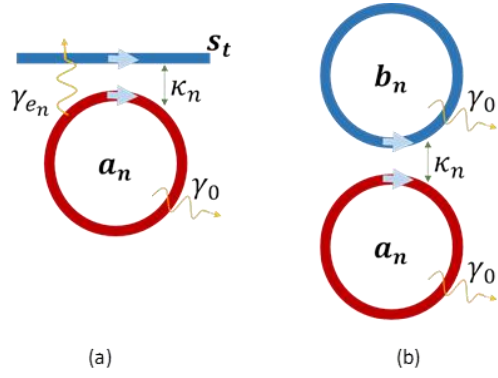

Figure S4. Schematics of the discussed mode management methods (a) The structure used for evanescent mode filtering of higher transverse modes, and (b) the PT-symmetric counterpart. Energy amplitudes are denoted by $a_n$ and $b_n$, the couplings as $\kappa_n$, $\gamma_0$ and $\gamma_{e_n}$ represent the linear loss due to scattering/absorption and loss due to coupling, respectively.

A comparison of the two cases, based on the numbers in Fig. 2 of the main text shows that $g_{max,PT}$ is at least an order of magnitude larger that $g_{max,EC}$. As an example, for the set of parameters (obtained from finite element simulations): $r_{avg} = 5.25$ μm, $n_{g_1} = 3.55$, $n_{g_0} = 4.09$, $\kappa_0 = 8.0 \times 10^{11}$ s$^{-1}$ and $\kappa_1 = 9.3 \times 10^{11}$ s$^{-1}$, one obtains $\Delta g_{max,PT}/\Delta g_{max,EC} = 19.1$.

**References**
S1. Little B.E., S. T. Chu, H. A. Haus, J. Foresi, and J. P. Laine, *J. Lightwave Technol.* **15**, 998 (1997).